\documentclass[onecolumn,preprint,showpacs,showkeys,superscriptaddress]{revtex4-1}
\usepackage{amssymb}
\usepackage{amsfonts}
\usepackage{amsmath}
\usepackage{graphicx}
\usepackage{epstopdf}
\usepackage{float}
\usepackage{relsize}

\begin{document}

\title{Klein--Gordon oscillator in a topologically nontrivial space-time}
\author{L. C. N. Santos}
\email{luis.santos@ufsc.br}
\affiliation{Departamento de F\'{\i}sica - CFM - Universidade Federal de Santa Catarina, CP. 476
- CEP 88.040 - 900, Florian\'{o}polis - SC - Brazil}
\author{C. E. Mota}
\email{clesio.evangelista@posgrad.ufsc.br}
\affiliation{Departamento de F\'{\i}sica - CFM - Universidade Federal de Santa Catarina, CP. 476
- CEP 88.040 - 900, Florian\'{o}polis - SC - Brazil}
\author{C. C. Barros Jr.}
\email{barros.celso@ufsc.br }
\affiliation{Departamento de F\'{\i}sica - CFM - Universidade Federal de Santa Catarina, CP. 476
- CEP 88.040 - 900, Florian\'{o}polis - SC - Brazil}
\begin{abstract}
In this study, we analyze solutions of the wave equation for scalar particles in a space-time with nontrivial topology. Solutions for the Klein--Gordon oscillator are found considering two configurations of this space-time. In the first one, it is assumed the $S^{1}\times R^{3}$ space where the metric is written in the usual inertial frame of reference. In the second case, we consider a rotating reference frame adapted to the circle $S^{1}$. We obtained compact expressions for the energy spectrum and for the particles wave functions in both configurations. Additionally, we show that the energy spectrum of the solution associated to the rotating system has an additional term that breaks the symmetry around $E = 0$.

\end{abstract}


\maketitle

\preprint{}

\affiliation{Departamento de F\'{\i}sica - CFM - Universidade Federal de
Santa Catarina, Florian\'{o}polis - SC - Brazil}

\affiliation{Departamento de F\'{\i}sica - CFM - Universidade Federal de Santa Catarina, CP. 476
- CEP 88.040 - 900, Florian\'{o}polis - SC - Brazil}

\volumeyear{} \volumenumber{} \issuenumber{} \eid{identifier} \startpage{1} %
\endpage{10}

\section{introduction}

\label{sec1}

The local and the global structures of space-time play an important role in the behavior of quantum systems. In this aspect, it is believed that global features of space-time may be directly related to the shift of energy levels of quantum particles. In the particular case of the $ S^{1}\times R^{1}$(time)$\times R^{2}$(space) space-time,  which
is locally flat but which has a nontrivial topology, one may consider the effect of periodic boundary conditions in one spatial direction. In this space-time we have one compactified spacelike dimension, thus it is expected that although the flat geometry, measurable effects occur in observable quantities. Nontrivial space-times have been studied extensively in literature, interesting applications are found in the study of atomic Bose-Einstein condensates  \cite{nontrivial1} with toroidal optical dipole traps. In the context of quantum field theory, vacuum polarization in a nonsimply connected space-time with the topology of $ S^{1}\times R^{3}$ is considered in \cite{nontrivial2}. It was
found that the vacuum energy for a free spinor field  in twisted and untwisted configurations are different in $ S^{1}\times R^{3}$ space. 

On the other hand, in quantum mechanics the harmonic oscillator is one of the most significant systems to be studied. In recent years, the relativistic version of the harmonic oscillator has been considered in several studies \cite{kgoscillator4,kgoscillator5,kgoscillator6,kgoscillator7,kgoscillator8,kgoscillator9,
kgoscillator10,diraco7,diraco8,diraco10,diraco11,diraco12,diraco13,santos5}. This important potential has been introduced as a linear interaction in the Klein--Gordon
equation \citep{kgoscillator1}. In the case of the Dirac equation, the so called Dirac oscillator has been introduced as an instance of a relativistic potential  such that its nonrelativistic limit leads to  the harmonic oscillator plus a strong spin-orbit coupling \cite{diraco9},  this result is similar to the one that is obtained for the  Klein--Gordon oscillator  when the spin-orbit is absent. Most recently, the relativistic harmonic oscillator has been studied in the context of the Kaluza--Klein theory  \citep{kgoscillator2}, where the Klein--Gordon oscillator coupled to a series of cosmic strings in five dimension  has been considered. In \cite{Castro1,Castro2} the author considers the effect of such kind of topological defect on scalar bosons described by the Duffin--Kemmer--Petiau
(DKP) formalism. The Klein--Gordon oscillator in a noncommutative phase space under a uniform magnetic field has been studied in \citep{kgoscillator3}. In this paper the authors conclude that the Klein--Gordon oscillator  in a noncommutative  space with an uniform magnetic field has behavior similar to the Landau problem in the usual space-time. 

 Another aspect of interest in our work is the influence of noninertial effects on quantum systems. As in classical physics, quantum mechanics is sensitive to the use of noninertial reference systems.  These effects can be taken into account through an appropriate coordinate transformation. Previous research reported in literature \cite{bakke2,bakke3} shows that rotating frames in the Minkowski space-time can play the role of a hard-wall potential. Recently these ideas have been applied to the case of spaces with nontrivial topology. In particular,  a rotating system was proposed recently in Ref. 
\cite{casimir28} where a scalar field on a circle (topology $S^{1}\times R^{1}$) with a Dirichlet cut has
been considered. In \cite{santos6},  a similar study was carried out in the case of a five-dimensional space-time. 

 Therefore, in this contribution, we will study bosons in the $ S^{1}\times R^{3}$  space-time by considering
the scalar wave equation for the Klein--Gordon oscillator. In fact, solutions of wave equations in curved spaces and nontrivial topology have been explored in various contexts \cite{santos1,santos2,santos5,santos6,adilberto1,reply1,reply2,reply3,reply4,string8,string12,string13,string14}. We will examine the combination of the  Klein--Gordon oscillator  and a space with nontrivial topology. Afterwards, a rotating frame in the $ S^{1}\times R^{3}$ space-time will be considered. We will show that the oscillator potential can form bound states for the Klein--Gordon equation in this space-time, and beyond that the momentum  associated with the nontrivial topology is discrete. This is an expected result, since the topology of $ S^{1}\times R^{3}$ space is associated with the periodicity of the boundary conditions. In the case of a rotating frame in the $ S^{1}\times R^{3}$ space-time, we will see interesting results associated with noninertial effects: the energy levels are shifted and the region of the space-time where the particle can be placed, is restricted. 

This work is organized as follows: In Section \ref{sec2}, we will study the space-time metric with a nontrivial topology and define a coordinate transformation that connects it to a rotating frame. In Section \ref{sec3}, we will derive the Klein--Gordon (KG) equation with a potential of the harmonic oscillator type and solve the associated differential equation. Similarly, we will solve again the KG equation in Section \ref{sec4} but we will consider a noninertial frame. Finally, we will present our conclusions in Section \ref{sec5}.
 
\section{Nontrivial space-time topology and noninertial reference frame}
\label{sec2}
\begin{figure}[H]
\includegraphics[scale=0.35]{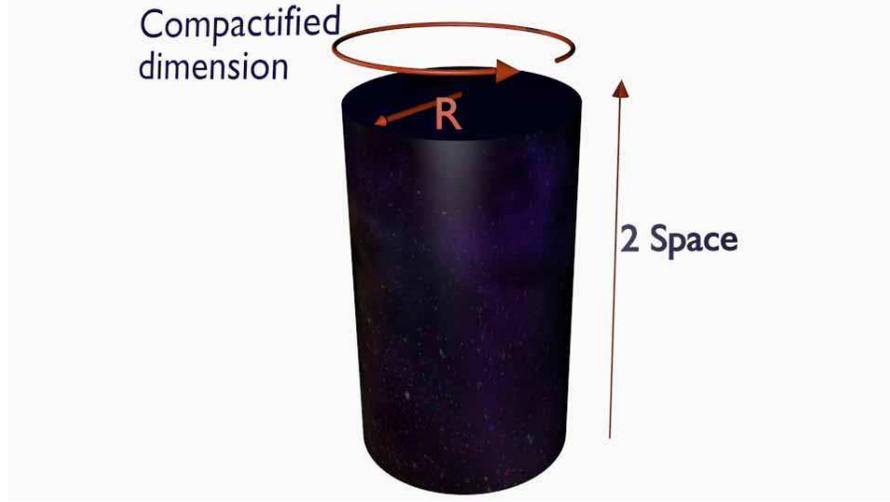}\newline
\caption{Representation of the topologically nontrivial space-time $S^{1}\times R^{3}$ . In this background, we discuss a frame with a constant angular velocity adapted to the circle $S^{1}$ }
\label{fig1}
\end{figure}

In this section, we define the line element that describes the space-time geometry in agreement with the proposal of this work. We want to study the behavior of massive scalar fields (zero spin particles) under the influence of a gravitational field generated by a space-time with the nontrivial topology $S^{1}\times R^{3}$. In this geometry, $R^{3}$ represents the usual uncompactified space-time directions, and $S^{1}$ is an extra compactified dimension. We discuss the relationship between $S^{1}\times R^{3}$ and the effects for a rotational frame inserted in that scenario.  Fig. \ref{fig1} shows a representation of this space-time where the temporal coordinate is absent.

The metric in polar coordinates describing the space-time under consideration is described by the expression

\begin{equation}
ds'^{2} = -dt'^{2}+dr'^{2}+r^{2}d\phi'^{2}+R^{2}d\theta'^{2},
\label{1}
\end{equation}
where, $-\infty < t' < \infty$ and $0<r'<\infty$ represent the temporal and radial coordinate range respectively. The parameter $R$ is the radius of the circle $S^{1}$, $\theta'$ and $\phi'$ are angular coordinates defined in the range $0 \leq \theta',\phi' \leq 2 \pi$. Also in this coordinate system, Eq. (\ref{1}) can be rewritten for a $S^{1}$ rotating system of reference with constant angular velocity $\Omega$, by means of the transformation
\begin{equation}
t' = t, \ r'=r, \ \phi' = \phi, \ \theta'=\theta + \Omega t,
\label{2}
\end{equation}
and inserting the coordinate transformation (\ref {2}) into Eq. (\ref {1}), we were able to get the line element
\begin{equation}\label{3}
\begin{split}
ds^{2} &= -dt^{2}+dr^{2}+r^{2}d\phi^{2}+R^{2}(d\theta +  \Omega dt )^{2} \\ 
&= -(1-R^{2}\Omega^{2})dt^{2} + dr^{2} + r^{2}d\phi^{2} + R^{2}d\theta^{2} + 2R^{2} \Omega d\theta dt, 
\end{split}
\end{equation}
where, we write the components of the covariant metric tensor related to Eq. (\ref{3}) in the matrix form

\begin{center}
$g_{\mu\nu}=$
$\begin{pmatrix}

  -(1-R^{2}\Omega^{2})  \ &  \  0  \ &  \ 0  \ & R^{2}\Omega \\
  
       0             & 1 & 0 & 0 \\
       
       0             & 0 &  \ r^{2} & 0 \\
       
      R^{2}\Omega    & 0 &  0    & R^{2} 

\end{pmatrix},$
\end{center}
together with its contravariant version

\begin{center}
$g^{\mu\nu}=$
$\begin{pmatrix}

  -1  \ &  \  0  \ &  \ 0  \ & \Omega \\
  
       0             & 1 & 0 & 0 \\
       
       0             & 0 &  \ \frac{1}{r^{2}} & 0 \\
       
     \Omega    & 0 &  0    & \frac{1}{R^{2}} - \Omega^{2} 

\end{pmatrix}.$
\end{center}

We can see that both $g_{\mu\nu}$ and $g^{\mu\nu}$ are non-diagonal. The non-null components outside the diagonal are effects originated from the rotational frame inserted in this scenario by means of the coordinate transformation (\ref{2}). Note that for $\Omega = 0$ we retrieve the line element described in Eq. (\ref{1}). From Eq. (\ref{3}), we can observe that $R$ is defined in the interval $0<R<R_{0}$ where $ R_{0} = \frac{1}{\Omega}$. That is, the rotating reference system can only be defined for distances where $R<\frac{1}{\Omega}$. In fact, since for $R>\frac{1}{\Omega}$, the time component $g_{00}$ of the metric becomes negative, so that such a condition is not physically acceptable. The non-compatibility of the rotational reference system with real particles at great distances is related to the fact that at that location the velocity of the frame becomes greater than the velocity of light in the vacuum and therefore such a system can not be defined from real bodies \cite{santos6}. Thus, it is convenient to restrict a range to $R$, where of course this condition imposes a size limit for the extra compact dimension in a noninertial frame. The direction of the movement occurs counterclockwise, that is, $\Omega\geq0$. 

\section{Klein--Gordon oscillator in nontrivial topology}
\label{sec3}
In this section we study the Klein--Gordon oscillator in a geometry $S^{1}\times R^{3}$. Spin-$0$ bosonic particles are described by a scalar field denoted as $\Psi(x)$. The dynamics of such a scalar field is determined by the so-called Klein--Gordon equation, to which we now turn our attention. Let us investigate, from this point on, the formulation of an equation equivalent to the scalar wave equation for the Klein--Gordon oscillator under the curved space-time under consideration and then to solve it. Let us first write in the Minkowski geometry. Recalling that a true real scalar field in this space-time (in units $c = \hbar = 1$) obeys the wave equation
\begin{equation}
(\eta^{\mu\nu}\partial_{\mu}\partial_{\nu}+m^{2})\Psi=0,
\end{equation}
where $m$ is the mass of the particle and $\eta^{\mu\nu}$ are the components of the metric tensor. The generalization of the Klein--Gordon equation for curved space-time, that is, a region affected by a gravitational field, is done by substituting the metric tensor $\eta^{\mu\nu}$ by the metric tensor that describes the curved space-time, $g^{\mu\nu}$, and the partial derivative  $\partial_{\mu}$ for the covariant derivative $\nabla_{\mu}$. Thus, the Klein--Gordon equation in a curved space-time is written as
\begin{equation}
\left[ \frac{1}{\sqrt{-g}}\partial_{\mu}(\sqrt{-g} g^{\mu\nu} \partial_{\nu})+ m^{2}\right] \Psi = 0,
\label{4}
\end{equation}  
where $g$ is the determinant of $g^{\mu\nu}$. Now, the quantum description of a particle takes into account elements of the space-time geometry in question. In this scenario, the coupling by means of a scalar interaction is given by the potential $ V(r)$, that is, an arbitrary scalar potential. This procedure is done by means of a redefinition of the mass term of the form: $m \rightarrow m + V(r)$. Substituting this redefinition for $ m$ in (\ref{4}) we obtain the following differential equation:
\begin{equation}
\left[-\frac{1}{\sqrt{-g}}\partial_{\mu}(g^{\mu\nu}\sqrt{-g}\partial_{\nu}) + (m + V)^{2}\right]\Psi=0.
\end{equation}

Let us now study the Klein--Gordon oscillator. Such a procedure is similar to one that is carried out for the insertion of an electromagnetic interaction which is made by the introduction of a 4-vector potential external vector $A_{\mu}$ (Gauge field). The Klein--Gordon oscillator at $S^{1}\times R^{3}$ is examined by the following transformation in the momentum operator $p_{\mu}$,
\begin{equation}
p_{\mu} \longrightarrow (p_{\mu} + im \omega X_{\mu}),
\end{equation} 
where $m$ is the mass of the particle, $\omega$ is the oscillation frequency and $X_{\mu}= (0,r,0,0) $ is the potential in polar coordinates in the radial direction $r$. Thus, the general form for the Klein--Gordon equation is given by equation
\begin{equation}
\left[ -\frac{1}{\sqrt{-g}}(\partial_{\mu}+m \omega X_{\mu})g^{\mu\nu} \sqrt{-g}(\partial_{\nu} - m \omega X_{\nu}) + (m+V)^{2} \right]\Psi = 0.
\label{5}
\end{equation}
Therefore, now we write Eq. (\ref{5}) in the space-time given by the line element (\ref{1}). Considering $V=0$, we obtain the equation:
\begin{equation}
\left[- \frac{\partial^{2}}{\partial t^{2}} + \frac{1}{r} \left( \frac{\partial}{\partial r} + m \omega r\right) \left( r \frac{\partial}{\partial r} - m \omega r^{2} \right) + \frac{1}{r^{2}} \frac{\partial^{2}}{\partial \phi^{2}}+ \frac{1}{R^{2}} \frac{\partial^{2}}{\partial \theta^{2}} - m^{2} \right]\Psi = 0,
\label{6}
\end{equation}
that may be written in the form
\begin{equation}
\left[-\frac{\partial^{2}}{\partial t^{2}} + (\frac{1}{r}\frac{\partial}{\partial r}) +        \frac{\partial^{2}}{\partial r^{2}} - 2m \omega - m^{2} \omega^{2} r^{2} + \frac{1}{r^{2}} \frac{\partial^{2}}{\partial \phi^{2}} + \frac{1}{R^{2}}\frac{\partial^{2}}{\partial \theta^{2}} - m^{2} \right]\Psi = 0,
\label{7} 
\end{equation}
that is independent of the coordinates $t$, $\theta$ and $\phi$. Thus, to solve Eq. (\ref{7}), let's assume that the solution is given as follows
\begin{equation}
\Psi(t,r,\theta,\phi) = f'(r)e^{-iEt}e^{+il\phi}e^{in\theta},
\label{8}
\end{equation}
where $l=n=0,\pm 1, \pm 2, \pm 3,$ are the quantum numbers, and $E$ is the energy of the particle. Substituting Eq. (\ref{8}) into Eq. (\ref{7}), we obtain
\begin{equation}
\left[\frac{d^{2}}{d r^{2}} + \frac{1}{r} \frac{d}{d r} - m^{2} \omega^{2} r^{2} - \frac{l^{2}}{r^{2}} + E^{2} - m^{2} - 2m \omega - \frac{n^{2}}{R^{2}}\right]f'(r) = 0,
\label{9}
\end{equation}
that is a second-order differential equation for the radial coordinate of the Klein--Gordon equation. Considering the transformation $f'(r) = \frac{F(r)}{\sqrt{r}}$ into Eq. (\ref{9}), we have the expression
\begin{equation}
\left[ \frac{d^{2}}{dr^{2}} - m^{2}\omega^{2}r^{2} - \frac{(l^{2}-\frac{1}{4})}{r^{2}} + \lambda^{2}\right] F(r)=0,
\label{10}
\end{equation}
with $ \lambda^{2}=E^{2} - m^{2} - 2m\omega - \frac{n^{2}}{R^{2}}$. The radial differential equation above describes the Klein--Gordon oscillator in a space-time with nontrivial topology. To obtain the solution of (\ref{10}), we first propose a transformation in the radial coordinate of the following form,
\begin{equation}
\rho = m\omega r^{2},
\label{11}
\end{equation}
which, inserted in the differential Eq. (\ref{10}), lead to the expression:
\begin{equation}
\left[\rho \frac{d^{2}}{d \rho^{2}}+ \frac{1}{2}\frac{d}{d\rho}-\frac{1}{4}\rho - \frac{(l^{2}-\frac{1}{4})}{4\rho}+\frac{\lambda^{2}}{4m\omega}\right]F(\rho)=0.
\label{12}
\end{equation}
Therefore, to get the normalizable eigenfunctions we can suppose a general solution in the  form 
\begin{equation}
F(\rho)=\rho^{\frac{1}{2}(\mid l \mid+ \frac{1}{2})}e^{\frac{-\rho}{2}}R(\rho),
\label{13}
\end{equation}
where, from Eq. (\ref{12}), by substituting expression (\ref{13}) for $F(\rho)$, and remembering that the parameter $l$ is a constant, we obtain the differential equation associated with the radial solution 
\begin{equation}
\rho \frac{d^{2}R(\rho)}{d\rho^{2}} + \left(\mid l \mid +1-\rho \right) \frac{d R(\rho)}{d\rho} 
- \left( \frac{1}{2}\mid l \mid + \frac{1}{2} - \frac{\lambda^{2}}{4 m \omega} \right) R(\rho) = 0. \label{14}
\end{equation}
We can observe that the final differential equation for $R(\rho)$ is of the confluent hypergeometric type or the Kummer equation, which is a second order that is linear homogeneous equation. Equation (\ref{14}) has as solution the hypergeometric function given by
\begin{equation}
R(\rho) =_{1} R_{1}(A,B;\rho),
\label{15b}
\end{equation}
where $A$ and $B$ are parameters defined by
\begin{equation}
A= \frac{1}{2}\mid l \mid + \frac{1}{2} - \frac{\lambda^{2}}{4 m \omega},
\label{16b}
\end{equation} 
\begin{equation}
B=\mid l \mid +1.
\label{17b}
\end{equation}
At this stage we may obtain the energy spectrum related to the confluent hypergeometric type solution. It is necessary that confluent hypergeometric function be a polynomial function of degree $N$ by considering its asymptotic behavior that demands that parameter $A$ be a negative integer \cite{abramo}.  Thus,
it is certainly possible to write the expression
\begin{equation}
A= \frac{1}{2}\mid l \mid + \frac{1}{2} - \frac{\lambda^{2}}{4 m \omega}=-N,\quad N=0,1,2,...
\label{18b}
\end{equation}

\begin{figure}[H]
\includegraphics[scale=0.9]{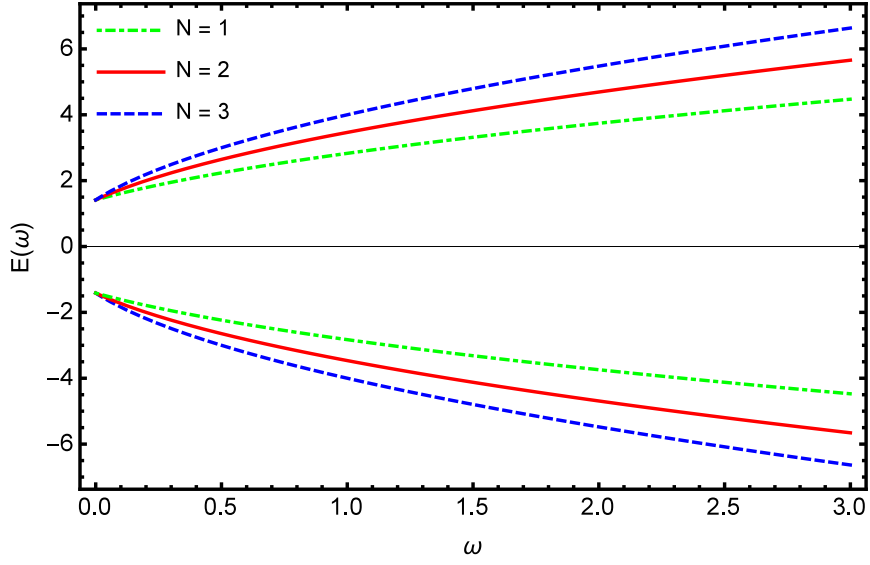}\newline
\caption{The plots of $E$ as functions of the variable $\omega$  displayed for three different values of $N$ with the parameters $m=1$, $R=1$, $n=1$.}
\label{fig2}
\end{figure}

Now we can solve this equation for energy, since $\lambda$ depends on $E$, the result is given by
\begin{equation}
E=\pm\sqrt{m^2+\frac{n^2}{R^2}+2m\omega(2N'+ \mid l\mid )}, \quad N'=N+1=1,2,3,...
\label{19b}
\end{equation}
It is interesting to note that the energy is symmetric around $E=0$, this means that particles and antiparticles have the same energy spectrum in this system. As shown in Fig. \ref{fig2}, it is possible to observe this behavior explicitly. In relation to the frequency of the oscillator, we can see that the energy increases, in absolute values, as $\omega$ grows. The spectrum also depends on the quantum number $n$ and the radius $R$ of the compactified dimension. Due to the mathematical form of expression (\ref{19b}), as $n$  increases, the values of the energy also increases.

\section{Klein--Gordon oscillator in noninertial reference frame}
\label{sec4}
In this section, we will study the influence of noninertial effects of a referential in rotation in a space-time with nontrivial topology, applied to the Klein--Gordon wave equation with a Klein--Gordon oscillator potential. Therefore, based on the procedures that were discussed in the previous section, in our computational developments that will be developed here, we will first recast the differential equation for the Klein--Gordon oscillator in the space-time described by the line element of Eq. (\ref{3}). Therefore, considering $V=0$, Eq. (\ref{5}) becomes
\begin{equation}
\left[ -\frac{\partial^{2}}{\partial t^{2}} + \frac{\partial^{2}}{\partial r^{2}} + \frac{1}{r}\frac{\partial}{\partial r} - 2m \omega -m^{2}\omega^{2}r^{2} + \frac{1}{r^{2}}\frac{\partial^{2}}{\partial \phi^{2}} + \left( \frac{1}{R^{2}} - \Omega^{2} \right) \frac{\partial^{2}}{\partial \theta^{2}} + 2\Omega \frac{\partial}{\partial t}\frac{\partial}{\partial \theta} - m^{2}\right] \Psi = 0. \label{15}
\end{equation}
This is the wave equation for spin-$0$ particles with the potential of the Klein--Gordon oscillator submitted to a rotating frame in a space-time with nontrivial topology. We can see that the solution of Eq. (\ref{15}) will have the same form discussed above, {\it i.e.}, (\ref{8}). Thus, by substituting (\ref{8}) into (\ref{15}), the expression is in effect,
\begin{equation}
\left[ \frac{d^{2}}{dr^{2}} + \frac{1}{r} \frac{d}{dr} - m^{2} \omega^{2} r^{2} - \frac{l^{2}}{r^{2}} + (E + \Omega n )^{2} - 2 m \omega - \frac{n^{2}}{R^{2}} - m^{2} \right] f'(r) =0.
\label{16}
\end{equation}
In this stage, by considering the transformation $ f'(r) = \frac{F(r)}{\sqrt{r}}$, we can write the radial equation (\ref{16}) in the form
\begin{equation}
\left[ \frac{d^{2}}{dr^{2}} - m^{2}\omega^{2}r^{2} - \frac{(l^{2}-\frac{1}{4})}{r^{2}} + \gamma^{2}\right] F(r)=0,
\label{17}
\end{equation}
where we define $\gamma^{2}=(E + \Omega n )^{2} - 2 m \omega - \frac{n^{2}}{R^{2}} - m^{2}$. Using the expression (\ref{11}), we can rewrite the radial Eq. (\ref{17}) in the following form,
\begin{equation}
\left[\rho \frac{d^{2}}{d \rho^{2}}+ \frac{1}{2}\frac{d}{d\rho}-\frac{1}{4}\rho - \frac{(l^{2}-\frac{1}{4})}{4\rho}+\frac{\gamma^{2}}{4m\omega}\right]F(\rho)=0,
\label{18}
\end{equation}
and therefore, by performing a substitution of the solution described in (\ref{13}) in the radial equation (\ref{18}), we find the expression
\begin{equation}
\rho \frac{d^{2}R(\rho)}{d\rho^{2}} + \left(\mid l \mid +1-\rho \right) \frac{d R(\rho)}{d\rho} 
- \left(\frac{1}{2}\mid l \mid + \frac{1}{2} - \frac{\gamma^{2}}{4 m \omega} \right) R(\rho) = 0. \label{19}
\end{equation}
The asymptotic behavior of hypergeometric confluent function implies that
\begin{equation}
\frac{1}{2}\mid l \mid + \frac{1}{2} - \frac{\gamma^{2}}{4 m \omega}=-N,\quad N=0,1,2,...
\label{20}
\end{equation}
Again, we can solve this equation for $E$ and the result obtained can be written in the form
\begin{equation}
E=-n\Omega\pm\sqrt{m^2+\frac{n^2}{R^2}+2m\omega(2N'+\mid l\mid)}, \quad N'=N+1=1,2,3,...,
\label{21}
\end{equation}
\begin{figure}[H]
\includegraphics[scale=0.9]{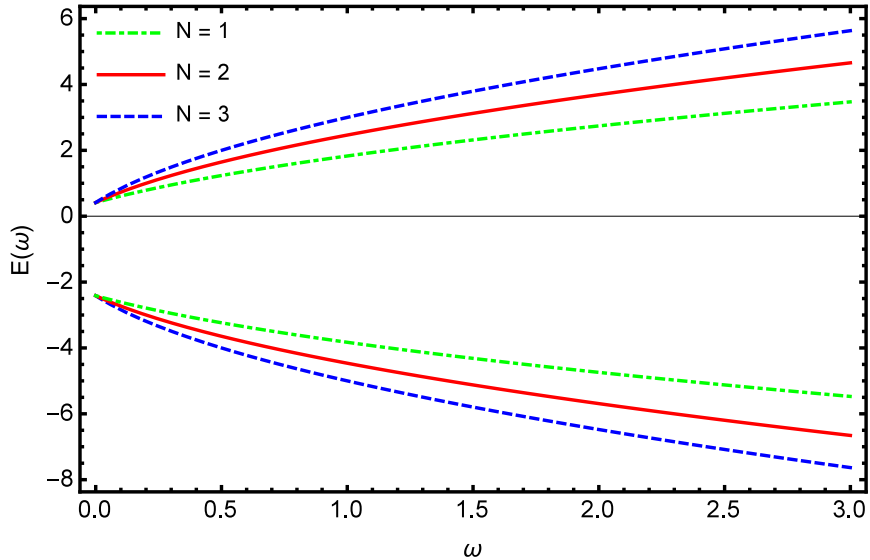}\newline
\caption{The plots of $E$ as functions of the variable $\omega$  displayed for three different values of $N$ with the parameters $m=1$, $R=1$, $n=1$, $\Omega=0.5$.}
\label{fig3}
\end{figure}
Now the energy spectrum depends on the angular velocity of the frame. The first term is related to the noninertial effects and appears frequently in this type of physical system. The second term is the usual energy spectrum of an inertial frame.
As it can be seen in Fig. \ref{fig3}, the effect of the referential rotation  breaks the symmetry around $E=0$.
\section{Conclusions}
\label{sec5}

In this work, we have determined solutions of the Klein--Gordon oscillator in a topologically nontrivial space-time. We have considered two different settings in this space. In the first case it is assumed the usual $S^{1}\times R^{3}$ space where the metric is written in the usual way and as a second case, it is considered a frame with a constant angular velocity  adapted in the circle $S^{1}$ by considering a coordinate  transformation. In both studied systems , we have solved the wave equations and have obtained the discrete energy spectrum associated with bound states. It was possible to see that the combined effect of a space with nontrivial topology and the Klein--Gordon oscillator allows the formation of bound states. When noninertial effects were taken into account, we verified that the energy spectrum lost the symmetry around $E=0$, i.e the additional term of the energy spectrum, in the case of the noninertial frame, causes a deviation from usual values. Other studies in the literature related to rotating reference systems show that the additional term that arises is associated with the coupling between the rotational angular momentum and the angular quantum number \cite{bakke2,bakke3,bakke4}. 

Additionally, we have shown that the space-time topology modifies the energy spectrum. In fact, the $S^{1}\times R^{3}$ space is associated with the periodicity of the boundary condition of the wave function. Consequently, the quantum number associated with the $S^{1}$ circle is discrete. In future works it may be interesting to extend the results obtained in this paper to spaces with different topologies and potentials. Effects of rotation on non-compact spatial coordinates is another type of configuration that we can study. In this way, the combination of inertial effects on different spatial coordinates can be useful in understanding the effect of the choice of noninertial reference frames in quantum mechanics.

\section{Acknowledgments}

This work was supported in part by means of funds provided by CAPES.

\bibliographystyle{aipnum4-1}
\bibliography{referencias_unificadas}

\begin{thebibliography}{39}%
\makeatletter
\providecommand \@ifxundefined [1]{%
 \@ifx{#1\undefined}
}%
\providecommand \@ifnum [1]{%
 \ifnum #1\expandafter \@firstoftwo
 \else \expandafter \@secondoftwo
 \fi
}%
\providecommand \@ifx [1]{%
 \ifx #1\expandafter \@firstoftwo
 \else \expandafter \@secondoftwo
 \fi
}%
\providecommand \natexlab [1]{#1}%
\providecommand \enquote  [1]{``#1''}%
\providecommand \bibnamefont  [1]{#1}%
\providecommand \bibfnamefont [1]{#1}%
\providecommand \citenamefont [1]{#1}%
\providecommand \href@noop [0]{\@secondoftwo}%
\providecommand \href [0]{\begingroup \@sanitize@url \@href}%
\providecommand \@href[1]{\@@startlink{#1}\@@href}%
\providecommand \@@href[1]{\endgroup#1\@@endlink}%
\providecommand \@sanitize@url [0]{\catcode `\\12\catcode `\$12\catcode
  `\&12\catcode `\#12\catcode `\^12\catcode `\_12\catcode `\%12\relax}%
\providecommand \@@startlink[1]{}%
\providecommand \@@endlink[0]{}%
\providecommand \url  [0]{\begingroup\@sanitize@url \@url }%
\providecommand \@url [1]{\endgroup\@href {#1}{\urlprefix }}%
\providecommand \urlprefix  [0]{URL }%
\providecommand \Eprint [0]{\href }%
\providecommand \doibase [0]{http://dx.doi.org/}%
\providecommand \selectlanguage [0]{\@gobble}%
\providecommand \bibinfo  [0]{\@secondoftwo}%
\providecommand \bibfield  [0]{\@secondoftwo}%
\providecommand \translation [1]{[#1]}%
\providecommand \BibitemOpen [0]{}%
\providecommand \bibitemStop [0]{}%
\providecommand \bibitemNoStop [0]{.\EOS\space}%
\providecommand \EOS [0]{\spacefactor3000\relax}%
\providecommand \BibitemShut  [1]{\csname bibitem#1\endcsname}%
\let\auto@bib@innerbib\@empty
%</preamble>
\bibitem [{\citenamefont {Wright}, \citenamefont {Arlt},\ and\ \citenamefont
  {Dholakia}(2000)}]{nontrivial1}%
  \BibitemOpen
  \bibfield  {author} {\bibinfo {author} {\bibfnamefont {E.}~\bibnamefont
  {Wright}}, \bibinfo {author} {\bibfnamefont {J.}~\bibnamefont {Arlt}}, \ and\
  \bibinfo {author} {\bibfnamefont {K.}~\bibnamefont {Dholakia}},\ }\href@noop
  {} {\bibfield  {journal} {\bibinfo  {journal} {Phys. Rev. A}\ }\textbf
  {\bibinfo {volume} {63}},\ \bibinfo {pages} {013608} (\bibinfo {year}
  {2000})}\BibitemShut {NoStop}%
\bibitem [{\citenamefont {Ford}(1980)}]{nontrivial2}%
  \BibitemOpen
  \bibfield  {author} {\bibinfo {author} {\bibfnamefont {L.}~\bibnamefont
  {Ford}},\ }\href@noop {} {\bibfield  {journal} {\bibinfo  {journal} {Physical
  Review D}\ }\textbf {\bibinfo {volume} {21}},\ \bibinfo {pages} {933}
  (\bibinfo {year} {1980})}\BibitemShut {NoStop}%
\bibitem [{\citenamefont {Jian-Hua}, \citenamefont {Kang},\ and\ \citenamefont
  {Sayipjamal}(2008)}]{kgoscillator4}%
  \BibitemOpen
  \bibfield  {author} {\bibinfo {author} {\bibfnamefont {W.}~\bibnamefont
  {Jian-Hua}}, \bibinfo {author} {\bibfnamefont {L.}~\bibnamefont {Kang}}, \
  and\ \bibinfo {author} {\bibfnamefont {D.}~\bibnamefont {Sayipjamal}},\
  }\href@noop {} {\bibfield  {journal} {\bibinfo  {journal} {Chin. Phys. C}\
  }\textbf {\bibinfo {volume} {32}},\ \bibinfo {pages} {803} (\bibinfo {year}
  {2008})}\BibitemShut {NoStop}%
\bibitem [{\citenamefont {Wen-Chao}(2003)}]{kgoscillator5}%
  \BibitemOpen
  \bibfield  {author} {\bibinfo {author} {\bibfnamefont {Q.}~\bibnamefont
  {Wen-Chao}},\ }\href@noop {} {\bibfield  {journal} {\bibinfo  {journal}
  {Chinese Phys.}\ }\textbf {\bibinfo {volume} {12}},\ \bibinfo {pages} {1054}
  (\bibinfo {year} {2003})}\BibitemShut {NoStop}%
\bibitem [{\citenamefont {Bakke}\ and\ \citenamefont
  {Furtado}(2015)}]{kgoscillator6}%
  \BibitemOpen
  \bibfield  {author} {\bibinfo {author} {\bibfnamefont {K.}~\bibnamefont
  {Bakke}}\ and\ \bibinfo {author} {\bibfnamefont {C.}~\bibnamefont
  {Furtado}},\ }\href {\doibase 10.1016/j.aop.2015.01.028} {\bibfield
  {journal} {\bibinfo  {journal} {Annals Phys.}\ }\textbf {\bibinfo {volume}
  {355}},\ \bibinfo {pages} {48} (\bibinfo {year} {2015})},\ \Eprint
  {http://arxiv.org/abs/1411.6988} {arXiv:1411.6988 [quant-ph]} \BibitemShut
  {NoStop}%
%%CITATION = ARXIV:1411.6988;%%
\bibitem [{\citenamefont {Maluf}(2011)}]{kgoscillator7}%
  \BibitemOpen
  \bibfield  {author} {\bibinfo {author} {\bibfnamefont {R.~V.}\ \bibnamefont
  {Maluf}},\ }\href {\doibase 10.1142/S0217751X11054887} {\bibfield  {journal}
  {\bibinfo  {journal} {Int. J. Mod. Phys. A}\ }\textbf {\bibinfo {volume}
  {26}},\ \bibinfo {pages} {4991} (\bibinfo {year} {2011})},\ \Eprint
  {http://arxiv.org/abs/1101.2801} {arXiv:1101.2801 [hep-th]} \BibitemShut
  {NoStop}%
\bibitem [{\citenamefont {Vit\'oria}, \citenamefont {Furtado},\ and\
  \citenamefont {Bakke}(2016)}]{kgoscillator8}%
  \BibitemOpen
  \bibfield  {author} {\bibinfo {author} {\bibfnamefont {R.~L.~L.}\
  \bibnamefont {Vit\'oria}}, \bibinfo {author} {\bibfnamefont {C.}~\bibnamefont
  {Furtado}}, \ and\ \bibinfo {author} {\bibfnamefont {K.}~\bibnamefont
  {Bakke}},\ }\href {\doibase 10.1016/j.aop.2016.03.016} {\bibfield  {journal}
  {\bibinfo  {journal} {Annals Phys.}\ }\textbf {\bibinfo {volume} {370}},\
  \bibinfo {pages} {128} (\bibinfo {year} {2016})},\ \Eprint
  {http://arxiv.org/abs/1511.05072} {arXiv:1511.05072 [quant-ph]} \BibitemShut
  {NoStop}%
%%CITATION = ARXIV:1511.05072;%%
\bibitem [{\citenamefont {Liang}\ and\ \citenamefont
  {Yang}(2012)}]{kgoscillator9}%
  \BibitemOpen
  \bibfield  {author} {\bibinfo {author} {\bibfnamefont {M.-L.}\ \bibnamefont
  {Liang}}\ and\ \bibinfo {author} {\bibfnamefont {R.-L.}\ \bibnamefont
  {Yang}},\ }\href {\doibase 10.1142/S0217751X12500479} {\bibfield  {journal}
  {\bibinfo  {journal} {Int. J. Mod. Phys. A}\ }\textbf {\bibinfo {volume}
  {27}},\ \bibinfo {pages} {1250047} (\bibinfo {year} {2012})}\BibitemShut
  {NoStop}%
\bibitem [{\citenamefont {Rao}\ and\ \citenamefont
  {Kagali}(2007)}]{kgoscillator10}%
  \BibitemOpen
  \bibfield  {author} {\bibinfo {author} {\bibfnamefont {N.~A.}\ \bibnamefont
  {Rao}}\ and\ \bibinfo {author} {\bibfnamefont {B.}~\bibnamefont {Kagali}},\
  }\href@noop {} {\bibfield  {journal} {\bibinfo  {journal} {Phys. Scripta}\
  }\textbf {\bibinfo {volume} {77}},\ \bibinfo {pages} {015003} (\bibinfo
  {year} {2007})}\BibitemShut {NoStop}%
\bibitem [{\citenamefont {Mandal}\ and\ \citenamefont {Verma}(2010)}]{diraco7}%
  \BibitemOpen
  \bibfield  {author} {\bibinfo {author} {\bibfnamefont {B.}~\bibnamefont
  {Mandal}}\ and\ \bibinfo {author} {\bibfnamefont {S.}~\bibnamefont {Verma}},\
  }\href@noop {} {\bibfield  {journal} {\bibinfo  {journal} {Phys. Lett. A}\
  }\textbf {\bibinfo {volume} {374}},\ \bibinfo {pages} {1021} (\bibinfo {year}
  {2010})}\BibitemShut {NoStop}%
\bibitem [{\citenamefont {Martinez-Y-Romero}, \citenamefont {Nunez-Yepez},\
  and\ \citenamefont {Salas-Brito}(1995)}]{diraco8}%
  \BibitemOpen
  \bibfield  {author} {\bibinfo {author} {\bibfnamefont {R.}~\bibnamefont
  {Martinez-Y-Romero}}, \bibinfo {author} {\bibfnamefont {H.}~\bibnamefont
  {Nunez-Yepez}}, \ and\ \bibinfo {author} {\bibfnamefont {A.}~\bibnamefont
  {Salas-Brito}},\ }\href@noop {} {\bibfield  {journal} {\bibinfo  {journal}
  {European Journal of Physics}\ }\textbf {\bibinfo {volume} {16}},\ \bibinfo
  {pages} {135} (\bibinfo {year} {1995})}\BibitemShut {NoStop}%
\bibitem [{\citenamefont {Pacheco}, \citenamefont {Landim},\ and\ \citenamefont
  {Almeida}(2003)}]{diraco10}%
  \BibitemOpen
  \bibfield  {author} {\bibinfo {author} {\bibfnamefont {M.}~\bibnamefont
  {Pacheco}}, \bibinfo {author} {\bibfnamefont {R.}~\bibnamefont {Landim}}, \
  and\ \bibinfo {author} {\bibfnamefont {C.}~\bibnamefont {Almeida}},\
  }\href@noop {} {\bibfield  {journal} {\bibinfo  {journal} {Phys. Lett. A}\
  }\textbf {\bibinfo {volume} {311}},\ \bibinfo {pages} {93} (\bibinfo {year}
  {2003})}\BibitemShut {NoStop}%
\bibitem [{\citenamefont {Quesne}\ and\ \citenamefont
  {Moshinsky}(1990)}]{diraco11}%
  \BibitemOpen
  \bibfield  {author} {\bibinfo {author} {\bibfnamefont {C.}~\bibnamefont
  {Quesne}}\ and\ \bibinfo {author} {\bibfnamefont {M.}~\bibnamefont
  {Moshinsky}},\ }\href@noop {} {\bibfield  {journal} {\bibinfo  {journal} {J.
  Phys. A: Math. Gen.}\ }\textbf {\bibinfo {volume} {23}},\ \bibinfo {pages}
  {2263} (\bibinfo {year} {1990})}\BibitemShut {NoStop}%
\bibitem [{\citenamefont {Rozmej}\ and\ \citenamefont
  {Arvieu}(1999)}]{diraco12}%
  \BibitemOpen
  \bibfield  {author} {\bibinfo {author} {\bibfnamefont {P.}~\bibnamefont
  {Rozmej}}\ and\ \bibinfo {author} {\bibfnamefont {R.}~\bibnamefont
  {Arvieu}},\ }\href@noop {} {\bibfield  {journal} {\bibinfo  {journal} {J.
  Phys. A: Math. Gen.}\ }\textbf {\bibinfo {volume} {32}},\ \bibinfo {pages}
  {5367} (\bibinfo {year} {1999})}\BibitemShut {NoStop}%
\bibitem [{\citenamefont {Villalba}(1994)}]{diraco13}%
  \BibitemOpen
  \bibfield  {author} {\bibinfo {author} {\bibfnamefont {V.}~\bibnamefont
  {Villalba}},\ }\href@noop {} {\bibfield  {journal} {\bibinfo  {journal}
  {Phys. Rev. A}\ }\textbf {\bibinfo {volume} {49}},\ \bibinfo {pages} {586}
  (\bibinfo {year} {1994})}\BibitemShut {NoStop}%
\bibitem [{\citenamefont {Santos}\ and\ \citenamefont
  {Barros}(2018{\natexlab{a}})}]{santos5}%
  \BibitemOpen
  \bibfield  {author} {\bibinfo {author} {\bibfnamefont {L.~C.~N.}\
  \bibnamefont {Santos}}\ and\ \bibinfo {author} {\bibfnamefont {C.~C.}\
  \bibnamefont {Barros}},\ }\href {\doibase 10.1140/epjc/s10052-017-5476-3}
  {\bibfield  {journal} {\bibinfo  {journal} {Eur. Phys. J. C}\ }\textbf
  {\bibinfo {volume} {78}},\ \bibinfo {pages} {13} (\bibinfo {year}
  {2018}{\natexlab{a}})}\BibitemShut {NoStop}%
\bibitem [{\citenamefont {Bruce}\ and\ \citenamefont
  {Minning}(1993)}]{kgoscillator1}%
  \BibitemOpen
  \bibfield  {author} {\bibinfo {author} {\bibfnamefont {S.}~\bibnamefont
  {Bruce}}\ and\ \bibinfo {author} {\bibfnamefont {P.}~\bibnamefont
  {Minning}},\ }\href@noop {} {\bibfield  {journal} {\bibinfo  {journal} {Il
  Nuovo Cimento A}\ }\textbf {\bibinfo {volume} {106}},\ \bibinfo {pages} {711}
  (\bibinfo {year} {1993})}\BibitemShut {NoStop}%
\bibitem [{\citenamefont {Moshinsky}\ and\ \citenamefont
  {Szczepaniak}(1989)}]{diraco9}%
  \BibitemOpen
  \bibfield  {author} {\bibinfo {author} {\bibfnamefont {M.}~\bibnamefont
  {Moshinsky}}\ and\ \bibinfo {author} {\bibfnamefont {A.}~\bibnamefont
  {Szczepaniak}},\ }\href@noop {} {\bibfield  {journal} {\bibinfo  {journal}
  {J. Phys. A: Math. Gen.}\ }\textbf {\bibinfo {volume} {22}},\ \bibinfo
  {pages} {L817} (\bibinfo {year} {1989})}\BibitemShut {NoStop}%
\bibitem [{\citenamefont {Carvalho}\ \emph {et~al.}(2016)\citenamefont
  {Carvalho}, \citenamefont {de~M.~Carvalho}, \citenamefont {Cavalcante},\ and\
  \citenamefont {Furtado}}]{kgoscillator2}%
  \BibitemOpen
  \bibfield  {author} {\bibinfo {author} {\bibfnamefont {J.}~\bibnamefont
  {Carvalho}}, \bibinfo {author} {\bibfnamefont {A.~M.}\ \bibnamefont
  {de~M.~Carvalho}}, \bibinfo {author} {\bibfnamefont {E.}~\bibnamefont
  {Cavalcante}}, \ and\ \bibinfo {author} {\bibfnamefont {C.}~\bibnamefont
  {Furtado}},\ }\href {\doibase 10.1140/epjc/s10052-016-4189-3} {\bibfield
  {journal} {\bibinfo  {journal} {Eur. Phys. J.}\ }\textbf {\bibinfo {volume}
  {C76}},\ \bibinfo {pages} {365} (\bibinfo {year} {2016})},\ \Eprint
  {http://arxiv.org/abs/1603.06292} {arXiv:1603.06292 [hep-th]} \BibitemShut
  {NoStop}%
\bibitem [{\citenamefont {Castro}(2016)}]{Castro1}%
  \BibitemOpen
  \bibfield  {author} {\bibinfo {author} {\bibfnamefont {L.~B.}\ \bibnamefont
  {Castro}},\ }\href {\doibase 10.1140/epjc/s10052-016-3904-4} {\bibfield
  {journal} {\bibinfo  {journal} {Eur. Phys. J. C}\ }\textbf {\bibinfo {volume}
  {76}},\ \bibinfo {pages} {1} (\bibinfo {year} {2016})}\BibitemShut {NoStop}%
\bibitem [{\citenamefont {Castro}(2015)}]{Castro2}%
  \BibitemOpen
  \bibfield  {author} {\bibinfo {author} {\bibfnamefont {L.~B.}\ \bibnamefont
  {Castro}},\ }\href {\doibase 10.1140/epjc/s10052-015-3507-5} {\bibfield
  {journal} {\bibinfo  {journal} {Eur. Phys. J. C}\ }\textbf {\bibinfo {volume}
  {75}},\ \bibinfo {pages} {1} (\bibinfo {year} {2015})}\BibitemShut {NoStop}%
\bibitem [{\citenamefont {Xiao}, \citenamefont {Long},\ and\ \citenamefont
  {Cai}(2011)}]{kgoscillator3}%
  \BibitemOpen
  \bibfield  {author} {\bibinfo {author} {\bibfnamefont {Y.-J.}\ \bibnamefont
  {Xiao}}, \bibinfo {author} {\bibfnamefont {Z.-W.}\ \bibnamefont {Long}}, \
  and\ \bibinfo {author} {\bibfnamefont {S.-H.}\ \bibnamefont {Cai}},\ }\href
  {\doibase 10.1007/s10773-011-0811-1} {\bibfield  {journal} {\bibinfo
  {journal} {Int. J. Theor. Phys.}\ }\textbf {\bibinfo {volume} {50}},\
  \bibinfo {pages} {3105} (\bibinfo {year} {2011})}\BibitemShut {NoStop}%
\bibitem [{\citenamefont {Bakke}(2010)}]{bakke2}%
  \BibitemOpen
  \bibfield  {author} {\bibinfo {author} {\bibfnamefont {K.}~\bibnamefont
  {Bakke}},\ }\href {\doibase http://dx.doi.org/10.1016/j.physleta.2010.09.046}
  {\bibfield  {journal} {\bibinfo  {journal} {Phys. Lett. A}\ }\textbf
  {\bibinfo {volume} {374}},\ \bibinfo {pages} {4642 } (\bibinfo {year}
  {2010})}\BibitemShut {NoStop}%
\bibitem [{\citenamefont {Bakke}(2013)}]{bakke3}%
  \BibitemOpen
  \bibfield  {author} {\bibinfo {author} {\bibfnamefont {K.}~\bibnamefont
  {Bakke}},\ }\href {\doibase 10.1142/S0217984913500188} {\bibfield  {journal}
  {\bibinfo  {journal} {Modern Physics Letters B}\ }\textbf {\bibinfo {volume}
  {27}},\ \bibinfo {pages} {1350018} (\bibinfo {year} {2013})}\BibitemShut
  {NoStop}%
\bibitem [{\citenamefont {Chernodub}(2013)}]{casimir28}%
  \BibitemOpen
  \bibfield  {author} {\bibinfo {author} {\bibfnamefont {M.~N.}\ \bibnamefont
  {Chernodub}},\ }\href {\doibase 10.1103/PhysRevD.87.025021} {\bibfield
  {journal} {\bibinfo  {journal} {Phys. Rev. D}\ }\textbf {\bibinfo {volume}
  {87}},\ \bibinfo {pages} {025021} (\bibinfo {year} {2013})},\ \Eprint
  {http://arxiv.org/abs/1207.3052} {arXiv:1207.3052 [quant-ph]} \BibitemShut
  {NoStop}%
%%CITATION = ARXIV:1207.3052;%%
\bibitem [{\citenamefont {Santos}\ and\ \citenamefont
  {Barros}(2018{\natexlab{b}})}]{santos6}%
  \BibitemOpen
  \bibfield  {author} {\bibinfo {author} {\bibfnamefont {L.~C.~N.}\
  \bibnamefont {Santos}}\ and\ \bibinfo {author} {\bibfnamefont {C.~C.}\
  \bibnamefont {Barros}},\ }\bibfield  {booktitle} {\emph {\bibinfo {booktitle}
  {Int. J. Mod. Phys. A}},\ }\href@noop {} {\ \textbf {\bibinfo {volume}
  {33}},\ \bibinfo {pages} {1850122} (\bibinfo {year}
  {2018}{\natexlab{b}})}\BibitemShut {NoStop}%
\bibitem [{\citenamefont {Santos}\ and\ \citenamefont
  {Barros}(2016)}]{santos1}%
  \BibitemOpen
  \bibfield  {author} {\bibinfo {author} {\bibfnamefont {L.~C.~N.}\
  \bibnamefont {Santos}}\ and\ \bibinfo {author} {\bibfnamefont {C.~C.}\
  \bibnamefont {Barros}},\ }\href@noop {} {\bibfield  {journal} {\bibinfo
  {journal} {Eur. Phys. J. C}\ }\textbf {\bibinfo {volume} {76}},\ \bibinfo
  {pages} {560} (\bibinfo {year} {2016})}\BibitemShut {NoStop}%
\bibitem [{\citenamefont {Santos}\ and\ \citenamefont
  {Barros}(2017)}]{santos2}%
  \BibitemOpen
  \bibfield  {author} {\bibinfo {author} {\bibfnamefont {L.~C.~N.}\
  \bibnamefont {Santos}}\ and\ \bibinfo {author} {\bibfnamefont {C.~C.}\
  \bibnamefont {Barros}},\ }\href {\doibase 10.1140/epjc/s10052-017-4732-x}
  {\bibfield  {journal} {\bibinfo  {journal} {Eur. Phys. J. C}\ }\textbf
  {\bibinfo {volume} {77}},\ \bibinfo {pages} {186} (\bibinfo {year}
  {2017})}\BibitemShut {NoStop}%
\bibitem [{\citenamefont {Andrade}, \citenamefont {Filgueiras},\ and\
  \citenamefont {Silva}(2017)}]{adilberto1}%
  \BibitemOpen
  \bibfield  {author} {\bibinfo {author} {\bibfnamefont {F.~M.}\ \bibnamefont
  {Andrade}}, \bibinfo {author} {\bibfnamefont {C.}~\bibnamefont {Filgueiras}},
  \ and\ \bibinfo {author} {\bibfnamefont {E.~O.}\ \bibnamefont {Silva}},\
  }\href@noop {} {\bibfield  {journal} {\bibinfo  {journal} {Adv. High Energy
  Phys.}\ }\textbf {\bibinfo {volume} {2017}},\ \bibinfo {pages} {8934691}
  (\bibinfo {year} {2017})},\ \Eprint {http://arxiv.org/abs/1604.05051}
  {1604.05051} \BibitemShut {NoStop}%
%%CITATION = ARXIV:1604.05051;%%
\bibitem [{\citenamefont {Vit\'oria}\ and\ \citenamefont
  {Bakke}(2018{\natexlab{a}})}]{reply1}%
  \BibitemOpen
  \bibfield  {author} {\bibinfo {author} {\bibfnamefont {R.~L.~L.}\
  \bibnamefont {Vit\'oria}}\ and\ \bibinfo {author} {\bibfnamefont
  {K.}~\bibnamefont {Bakke}},\ }\href {\doibase 10.1140/epjc/s10052-018-5658-7}
  {\bibfield  {journal} {\bibinfo  {journal} {Eur. Phys. J. C}\ }\textbf
  {\bibinfo {volume} {78}},\ \bibinfo {pages} {175} (\bibinfo {year}
  {2018}{\natexlab{a}})},\ \Eprint {http://arxiv.org/abs/1802.07536}
  {arXiv:1802.07536 [gr-qc]} \BibitemShut {NoStop}%
%%CITATION = ARXIV:1802.07536;%%
\bibitem [{\citenamefont {Ahmed}(2018)}]{reply2}%
  \BibitemOpen
  \bibfield  {author} {\bibinfo {author} {\bibfnamefont {F.}~\bibnamefont
  {Ahmed}},\ }\href {\doibase 10.1140/epjc/s10052-018-6082-8} {\bibfield
  {journal} {\bibinfo  {journal} {Eur. Phys. J. C}\ }\textbf {\bibinfo {volume}
  {78}},\ \bibinfo {pages} {598} (\bibinfo {year} {2018})}\BibitemShut
  {NoStop}%
\bibitem [{\citenamefont {Vit\'oria}\ and\ \citenamefont
  {Bakke}(2018{\natexlab{b}})}]{reply3}%
  \BibitemOpen
  \bibfield  {author} {\bibinfo {author} {\bibfnamefont {R.~L.~L.}\
  \bibnamefont {Vit\'oria}}\ and\ \bibinfo {author} {\bibfnamefont
  {K.}~\bibnamefont {Bakke}},\ }\href {\doibase 10.1140/epjp/i2018-12310-9}
  {\bibfield  {journal} {\bibinfo  {journal} {Eur. Phys. J. Plus}\ }\textbf
  {\bibinfo {volume} {133}},\ \bibinfo {pages} {490} (\bibinfo {year}
  {2018}{\natexlab{b}})}\BibitemShut {NoStop}%
%%CITATION = EPHJP,133,490;%%
\bibitem [{\citenamefont {Wang}\ \emph {et~al.}(2018)\citenamefont {Wang},
  \citenamefont {Long}, \citenamefont {Long},\ and\ \citenamefont
  {Wu}}]{reply4}%
  \BibitemOpen
  \bibfield  {author} {\bibinfo {author} {\bibfnamefont {B.-Q.}\ \bibnamefont
  {Wang}}, \bibinfo {author} {\bibfnamefont {Z.-W.}\ \bibnamefont {Long}},
  \bibinfo {author} {\bibfnamefont {C.-Y.}\ \bibnamefont {Long}}, \ and\
  \bibinfo {author} {\bibfnamefont {S.-R.}\ \bibnamefont {Wu}},\ }\href
  {\doibase 10.1142/S0217751X18501580} {\bibfield  {journal} {\bibinfo
  {journal} {Int. J. Mod. Phys. A}\ }\textbf {\bibinfo {volume} {33}},\
  \bibinfo {pages} {1850158} (\bibinfo {year} {2018})}\BibitemShut {NoStop}%
%%CITATION = IMPAE,A33,1850158;%%
\bibitem [{\citenamefont {Carvalho}, \citenamefont {de~M.~Carvalho},\ and\
  \citenamefont {Furtado}(2014)}]{string8}%
  \BibitemOpen
  \bibfield  {author} {\bibinfo {author} {\bibfnamefont {J.}~\bibnamefont
  {Carvalho}}, \bibinfo {author} {\bibfnamefont {A.}~\bibnamefont
  {de~M.~Carvalho}}, \ and\ \bibinfo {author} {\bibfnamefont {C.}~\bibnamefont
  {Furtado}},\ }\href {\doibase 10.1140/epjc/s10052-014-2935-y} {\bibfield
  {journal} {\bibinfo  {journal} {Eur. Phys. J. C}\ }\textbf {\bibinfo {volume}
  {74}} (\bibinfo {year} {2014}),\ 10.1140/epjc/s10052-014-2935-y}\BibitemShut
  {NoStop}%
\bibitem [{\citenamefont {Cavalcanti~de Oliveira}\ and\ \citenamefont
  {Bezerra~de Mello}(2006)}]{string12}%
  \BibitemOpen
  \bibfield  {author} {\bibinfo {author} {\bibfnamefont {A.~L.}\ \bibnamefont
  {Cavalcanti~de Oliveira}}\ and\ \bibinfo {author} {\bibfnamefont {E.~R.}\
  \bibnamefont {Bezerra~de Mello}},\ }\href {\doibase
  10.1088/0264-9381/23/17/009} {\bibfield  {journal} {\bibinfo  {journal}
  {Class. Quant. Grav.}\ }\textbf {\bibinfo {volume} {23}},\ \bibinfo {pages}
  {5249} (\bibinfo {year} {2006})},\ \Eprint
  {http://arxiv.org/abs/hep-th/0603036} {arXiv:hep-th/0603036 [hep-th]}
  \BibitemShut {NoStop}%
\bibitem [{\citenamefont {Figueiredo~Medeiros}\ and\ \citenamefont
  {de~Mello}(2012)}]{string13}%
  \BibitemOpen
  \bibfield  {author} {\bibinfo {author} {\bibfnamefont {E.~R.}\ \bibnamefont
  {Figueiredo~Medeiros}}\ and\ \bibinfo {author} {\bibfnamefont {E.~R.~B.}\
  \bibnamefont {de~Mello}},\ }\href {\doibase 10.1140/epjc/s10052-012-2051-9}
  {\bibfield  {journal} {\bibinfo  {journal} {Eur. Phys. J.}\ }\textbf
  {\bibinfo {volume} {C72}},\ \bibinfo {pages} {2051} (\bibinfo {year}
  {2012})},\ \Eprint {http://arxiv.org/abs/1108.3786} {arXiv:1108.3786
  [hep-th]} \BibitemShut {NoStop}%
\bibitem [{\citenamefont {Hosseinpour}\ \emph {et~al.}(2017)\citenamefont
  {Hosseinpour}, \citenamefont {Andrade}, \citenamefont {Silva},\ and\
  \citenamefont {Hassanabadi}}]{string14}%
  \BibitemOpen
  \bibfield  {author} {\bibinfo {author} {\bibfnamefont {M.}~\bibnamefont
  {Hosseinpour}}, \bibinfo {author} {\bibfnamefont {F.~M.}\ \bibnamefont
  {Andrade}}, \bibinfo {author} {\bibfnamefont {E.~O.}\ \bibnamefont {Silva}},
  \ and\ \bibinfo {author} {\bibfnamefont {H.}~\bibnamefont {Hassanabadi}},\
  }\href {https://doi.org/10.1140/epjc/s10052-017-4834-5} {\bibfield  {journal}
  {\bibinfo  {journal} {Eur. Phys. J. C}\ }\textbf {\bibinfo {volume} {77}},\
  \bibinfo {pages} {270} (\bibinfo {year} {2017})}\BibitemShut {NoStop}%
\bibitem [{\citenamefont {Abramowitz}\ and\ \citenamefont
  {Stegun}(1964)}]{abramo}%
  \BibitemOpen
  \bibfield  {author} {\bibinfo {author} {\bibfnamefont {M.}~\bibnamefont
  {Abramowitz}}\ and\ \bibinfo {author} {\bibfnamefont {I.}~\bibnamefont
  {Stegun}},\ }\href {https://books.google.com.br/books?id=MtU8uP7XMvoC} {\emph
  {\bibinfo {title} {Handbook of Mathematical Functions: With Formulas, Graphs,
  and Mathematical Tables}}},\ Applied mathematics series\ (\bibinfo
  {publisher} {Dover Publications},\ \bibinfo {year} {1964})\BibitemShut
  {NoStop}%
\bibitem [{\citenamefont {Bakke}\ and\ \citenamefont {Furtado}(2009)}]{bakke4}%
  \BibitemOpen
  \bibfield  {author} {\bibinfo {author} {\bibfnamefont {K.}~\bibnamefont
  {Bakke}}\ and\ \bibinfo {author} {\bibfnamefont {C.}~\bibnamefont
  {Furtado}},\ }\href {\doibase 10.1103/PhysRevD.80.024033} {\bibfield
  {journal} {\bibinfo  {journal} {Phys. Rev. D}\ }\textbf {\bibinfo {volume}
  {80}},\ \bibinfo {pages} {024033} (\bibinfo {year} {2009})}\BibitemShut
  {NoStop}%
\end{thebibliography}%

\end{document}